\documentclass[review]{elsarticle}

\usepackage{lineno,hyperref}
\usepackage{color}
\usepackage{amsmath,amsfonts,amssymb,esvect,ulem}
\usepackage{float}
\usepackage{morefloats}
\usepackage{graphicx}
\usepackage{placeins}
\usepackage{comment}

\usepackage{epstopdf}
\usepackage{float}
\usepackage{caption}
\usepackage{subcaption}
\modulolinenumbers[5]

\newcommand{\TAMU}{Texas A\&M University}

\begin{document}

\begin{frontmatter}

\title{Adjoint-based Sensitivity Analysis for High-Energy Density Radiative Transfer using Flux-Limited Diffusion }

\author[mymainaddress]{Kelli D. Humbird}
\ead{khumbird@tamu.edu}

\author[mymainaddress]{Ryan G. McClarren\corref{mycorrespondingauthor}}
\cortext[mycorrespondingauthor]{Corresponding author}
\ead{rgm@tamu.edu}

\address[mymainaddress]{Nuclear Engineering, \TAMU,~College Station, TX 77843-3133}

\begin{abstract}
Uncertainty quantification and sensitivity analyses are a vital component for predictive modeling in the sciences and engineering. The adjoint approach to sensitivity analysis requires solving a primary system of equations and a mathematically related set of adjoint equations. The information contained in the equations can be combined to produce sensitivity information in a computationally efficient manner. In this work, sensitivity analyses are performed on systems described by flux-limited radiative diffusion using the adjoint approach. The sensitivities computed are shown to agree with standard perturbation theory and require significantly less computational time. The adjoint approach saves the computational cost of one forward solve per sensitivity, making the method attractive when multiple sensitivities are of interest. 
\end{abstract}


\end{frontmatter}

\linenumbers

\section{Introduction}
The adjoint approach to performing sensitivity analyses is an efficient method for identifying parameters that have the greatest influence on a particular quantity of interest (QOI). The adjoint method requires formulation of a second problem, mathematically related to the forward system of equations, and uses the solution to obtain sensitivity information. This approach allows multiple sensitivities to be computed by solving the forward and adjoint equations once, and evaluating inner products for each sensitivity. This is computationally efficient compared to the perturbation approach to finding sensitivities, which requires solving the forward problem twice per perturbed parameter. The sensitivity is then found by dividing the change in the QOI by the change in the perturbed parameter. \\
The main disadvantage of the adjoint approach is that it can quickly become memory intensive ~\cite{Restrepo},~\cite{Sandu2}. The forward and adjoint solutions, along with parameter values, must be stored for all time, and at all spatial locations, in order to compute the sensitivities. For high fidelity, transient calculations the memory demand can rapidly exceed that available on standard computers. Typically, writing and reading to files is required, or a subset of information is stored and the solutions are recomputed from these checkpoints \cite{Stripling2013}. \\
The methodology for deriving an adjoint system of coupled differential equations is well documented~\cite{Cao2003}, \cite{Hartmann}, \cite{Sandu}. In this work, the adjoint equations are derived for a system of coupled partial differential equations describing radiative flux-limited diffusion. Approximations to complex expressions that result from the non-linear flux-limited diffusion model are shown not to introduce significant error when computing sensitivities. The resulting system of adjoint equations are linear; for the considered examples the adjoint sensitivity analysis takes less computational time than is required to compute a single sensitivity using the perturbation method.

\section{Adjoint-based sensitivity analysis}

An adjoint-based sensitivity analysis is performed on a system described by flux-limited radiative diffusion with material temperature feedback. The evolution of the forward system is described by a set of coupled partial differential equations~\cite{Brown}, \cite{MC2011}, \cite{Bowers}:
\begin{equation}\label{1} 
\begin{split}
\frac{1}{c}\frac{\partial \phi}{\partial t} -\nabla \cdot D \nabla \phi = S-\kappa\rho (\phi -acT^4), \\
\rho C_\mathrm{v} \frac{\partial T}{\partial t} = \kappa\rho \left(\phi-acT^4 \right),
\end{split}
\end{equation}
where $\phi$=$\phi$$(r,t)$ is the scalar intensity with units of GJ/cm$^2$ns, $T=T(r,t)$ is the temperature in keV, $c$ is 29.98 cm/ns, $a$ is 0.01372 GJ/cm$^3$-keV$^4$ , $S$ is an external volumetric source, $\kappa$ is the opacity, $\rho$ is the density, $C$$_\mathrm{v}$ is the specific heat, and $D$ is the flux-limited diffusion coefficient. Note that the solutions, $\phi$ and $T$, depend implicitly on the material constants and the diffusion coefficient. \\

Flux-limited diffusion coefficients are designed to correct for non-physical results related to the speed of propagation of information. In the diffusion approximation, the current is given by the expression:

\begin{equation}
J=-D\nabla\phi.
\end{equation}

The approximation is inaccurate where the gradient of the scalar intensity is large; flux-limited diffusion coefficients prevent the current from exceeding physical values in such regions. The Larsen
coefficient is used in this work~\cite{Hall2000}:
\begin{equation} \label{eq:d}
D=\left( (3\kappa\rho)^n+\left(\frac{1}{\phi}\left| \nabla \phi\right|\right)^{n} \right)^{-1/n}.
\end{equation}
When the gradient of the scalar intensity is small, this reduces to the standard diffusion coefficient; when the gradient is large, the second term in the denominator dominates, thus preventing the current from exceeding the value of the scalar intensity. Typically $n$ is chosen to be two, but it can be adjusted such that the diffusion solution agrees more closely with transport calculations~\cite{McClarren2012}.

To derive the equations for the adjoint scalar intensity, $\phi$$^{\dag}$, and the adjoint temperature, $T$$^{\dag}$, we use the Lagrangian approach to form the sensitivity expression. First, the differential equations are combined to form the operator $F$:
\begin{equation}\label{2} 
F=\binom{F_1}{F_2}=\binom{\dot{\phi}/c -f(\phi,T)}{\dot{T}-g(\phi,T)}=0,
\end{equation}
with
\begin{equation}\label{3} 
\begin{split}
f(\phi,T)=\nabla \cdot D \nabla \phi-\kappa\rho (\phi-acT^4) + S,\\
g(\phi,T)=\frac{\kappa\rho}{\rho C_\mathrm{v}}(\phi-acT^4).
\end{split}
\end{equation}

Following the work of Stripling \cite{Stripling2013} and Stripling, Anitescu, and Adams \cite{Stripling2012}, to find the adjoint system for the coupled set of equations, a Lagrangian is formed:
\begin{equation}\label{4} 
\mathcal{L}=\int \left[ \langle Q \rangle-\langle \lambda,F \rangle   \right] dt,
\end{equation}
where the angular brackets denote integrals over all space and $\langle$$Q$$\rangle$ is the quantity of interest. The integral over time is taken from $t$$_0$ to $t$$_f$.
The operator $F$ is defined to be zero, thus the sensitivities of $\langle$$Q$$\rangle$ and $\mathcal{L}$ are equivalent. 
To derive the expression for the sensitivity, the functional derivative of the Lagrangian is taken with respect to $\theta$ using the chain rule:
\begin{equation}\label{5}
\frac{\partial\mathcal{L}}{\partial\theta}= \int\left[\langle Q \rangle_{\theta}+\langle Q \rangle_{x}x_{\theta}-\frac{\partial}{\partial \dot{x}} \langle \lambda,F \rangle \dot{x}_{\theta}- \frac{\partial}{\partial x}\langle \lambda,F \rangle x_{\theta}- \frac{\partial}{\partial\theta}\langle \lambda,F \rangle \right] dt.
\end{equation}
In this equation, $x=($$\phi$ $T)$$^{T}$,
the subscripts denote partial derivatives, $\dot{x}$ is the time derivative of $x$, and $\lambda$ is an undetermined two component Lagrange multiplier. 
Using integration by parts, the integral can be rewritten as:
\begin{equation}\label{6}
\begin{split}
\frac{\partial \mathcal{L}}{\partial\theta}=\left[-\frac{\partial}{\partial\dot{x}}\langle\lambda, F \rangle x_{\theta} \right]_{t_{0}}^{t_{f}} + \int\left[\langle Q_{\theta} \rangle-\frac{\partial}{\partial\theta} \langle \lambda, F \rangle \right] dt \\
+ \int\left[ \langle Q_{x} \rangle + \frac{d}{dt}\left(\frac{\partial}{\partial\dot{x}}\langle\lambda, F\rangle \right)
-\frac{\partial}{\partial x}\langle\lambda, F\rangle \right] x_{\theta} \, dt,
\end{split}
\end{equation}
The only term in the above expression that cannot be computed directly is $x$$_{\theta}$, the derivative of the solution vector with respect to $\theta$. If this term was known, the adjoint approach would not be necessary as the sensitivity could be computed by direct differentiation of the QOI. 

The adjoint equations are defined to be the conditions that are imposed on $\lambda$$\equiv$$($$\phi$$^{\dag}$ $T$$^{\dag}$$ )$$^{T}$ such that the integrand of the final term is eliminated. Thus, the adjoint equations are given by the expression:
\begin{equation}\label{7}
\begin{split}
\left[-\frac{\partial}{\partial\dot{x}}\langle\lambda, F \rangle x_{\theta} \right]_{t_{0}} +  \langle Q_{x} \rangle + \frac{d}{dt}\left(\frac{\partial}{\partial\dot{x}}\langle\lambda, F\rangle \right)
-\frac{\partial}{\partial x}\langle\lambda, F\rangle=0.
\end{split}
\end{equation}
Note that the first term in \eqref{7} includes the sensitivity of the initial conditions to the parameter $\theta$. This term can be included if known; in the work that follows it is simply assumed that the sensitivity of the initial conditions is not of interest and the term is taken to be zero; this assumption is not necessary in general.
The value of $x$$_{\theta}$ at the final time is not known, and must be eliminated by imposing appropriate terminal conditions on the adjoint variables. The resulting system of adjoint equations evolve backward in time, with the terminal condition:
\begin{equation}\label{17}
\lambda(t_f)=0.
\end{equation}
Physically, the terminal condition states that events occurring beyond the final time step do not influence the adjoint solution ~\cite{Stripling2013},~\cite{Glasstone}.
 
To summarize, the forward system of equations (Eq. \eqref{1}) are solved provided initial conditions and the adjoint system of equations (Eq. \eqref{7}) are solved by imposing terminal conditions (Eq. \eqref{17}). The forward and adjoint solutions can then be used to evaluate the sensitivity of the QOI with respect to any parameter, $\theta$, by evaluating Eq. \eqref{12}:
\begin{equation} \label{12}
\frac{\partial \mathcal{L}}{\partial\theta} = \int\left[\langle Q_{\theta} \rangle-\frac{\partial}{\partial\theta} \langle \lambda, F \rangle \right] dt .
\end{equation}

The QOI for the following examples is the time integrated absorption, $\mathcal{A}$, in particular regions, thus:
\begin{equation}
\langle Q \rangle = \int \rho\kappa\phi \,\, dV,
\end{equation}
where the spatial integral is taken over the volume of interest. 
 The expressions for $F$ and $\langle$$Q$$\rangle$ are substituted into equation~\eqref{7}, and the system is split into its constituent components. The result is a set of coupled linear partial differential equations for the adjoint scalar intensity and the adjoint temperature:
\begin{equation}\label{8} 
\begin{split}
-\frac{1}{c}\frac{\partial \phi^\dag}{\partial t}-\frac{\partial}{\partial\phi}\nabla\cdot(D \nabla \phi)\phi^\dag+\kappa\rho \phi^\dag = \kappa\rho + \frac{\kappa}{C_\mathrm{v}} T^\dag, \\
-\frac{\partial T^\dag}{\partial t}=-4ac \kappa\rho T^3 \left(\frac{T^\dag}{\rho C_\mathrm{v}}-\phi^\dag\right) +\left(\nabla\frac{\partial D}{\partial T}\nabla\phi-\frac{\partial (\kappa\rho)}{\partial T}(\phi-acT^4) \right)\phi^\dagger \\
+\frac{T^\dag}{\rho C_\mathrm{v}}\left(\phi-acT^4 \right)\frac{\partial(\kappa\rho)}{\partial T}+ \phi\frac{\partial(\kappa\rho)}{\partial T}
.
\end{split}
\end{equation}

Equation \eqref{8} allows for temperature dependence in $D$ and $\kappa\rho$. The final expression for the sensitivity of the absorption, $\mathcal{A}$, with respect to an arbitrary parameter $\theta$ is given by:
\begin{equation}\label{9}
\begin{split}
\frac{\partial \mathcal{A} }{\partial\theta}=\left[\frac{\partial}{\partial\dot{x}}\langle\lambda, F \rangle x_{\theta} \right]_{t_{0}} + \int\left[\langle Q_{\theta} \rangle-\frac{\partial}{\partial\theta} \langle \lambda, F \rangle \right] dt .
\end{split}
\end{equation}

\subsection{Sensitivity Examples}
In this study the primary quantity of interest is the total absorption within particular volumes:
\begin{align*}
\mathcal{A}=\int\!\int\! \kappa\rho \, \phi \,\,\, dV \,dt,
\end{align*}
 where the integration over time is taken from $t$$_0$ to $t$$_f$ and the spatial integration is taken over a region of interest. When computing sensitivities, a few complicated derivatives must be evaluated. 
 For example, when computing the sensitivity of the absorption with respect to the opacity, the following derivative appears in the final term of equation ~\eqref{9}:
\begin{equation}\label{10}
\frac{\partial}{\partial\kappa}\left(-\nabla\cdot\left(\left\{ (3\kappa\rho)^n+\left(\frac{1}{\phi}\left|\nabla \phi\right|\right)^{n} \right\}^{-1/n}\nabla \phi \right) \right).
\end{equation}

The flux-limited form of the diffusion coefficient complicates the spatial derivative. To avoid performing this derivative analytically, a second-order centered finite difference approximation is employed. For a particular direction in Cartesian geometry, the expression above is estimated by:
\begin{equation}\label{11}
-\frac{\partial}{\partial\kappa}\left(\frac{\partial}{\partial r}D\frac{\partial \phi}{\partial r} \right)_{i} \approx -\frac{1}{\Delta r} \left (\left(\frac{\partial D}{\partial \kappa}\frac{\partial \phi}{\partial r} \right)_{i+1/2}-\left(\frac{\partial D}{\partial \kappa}\frac{\partial \phi}{\partial r} \right)_{i-1/2} \right),
\end{equation}
where $\Delta$$r$ is the discretization step size for coordinate $r$ and $i$$\pm$$1/2$ subscripts denote where on the discretized domain the functions should be evaluated. The derivative of the diffusion coefficient with respect to the opacity is calculated analytically. Similar approximations are made for the sensitivity of the absorption with respect to the parameter $n$ in the diffusion coefficient.

\subsection{Numerical Methods}
The adjoint equations are derived in their continuous form, however in practice they are solved numerically. The discretization scheme must be chosen such that the properties of the adjoint equations are preserved. Many Runge-Kutta schemes have been shown to retain the properties of the continuous adjoint equations, such as the implicit Euler and fourth order Runge-Kutta discretizations~\cite{Stripling2013}. Furthermore, these schemes possess symmetries that allow for integration forward and backward in time without requiring modification to the algorithm. For the forward equations an initial condition is supplied; for the adjoint equations a terminal condition is specified. \\
In this work, the implicit Euler discretization is used to solve the forward and adjoint systems of equations. The equation for the radiation scalar intensity becomes nonlinear with use of the flux-limited diffusion coefficient. In order to define a unique adjoint operator, the diffusion coefficient is linearized by lagging $\phi$ one time step in the forward solve, and using the known forward scalar intensity in the expression for $D$ to solve the adjoint equations.\\
The nonlinear forward temperature equation is also solved implicitly; the solution is converged when the relative L-2 norm of the solutions $\phi$ and $T$ changes less than 10$^{-8}$ between iterations. The adjoint system is solved using similar convergence criteria.

\section{Crooked Pipe Example}
\begin{figure}[H] 
	\centering
	\includegraphics[width=1\textwidth]{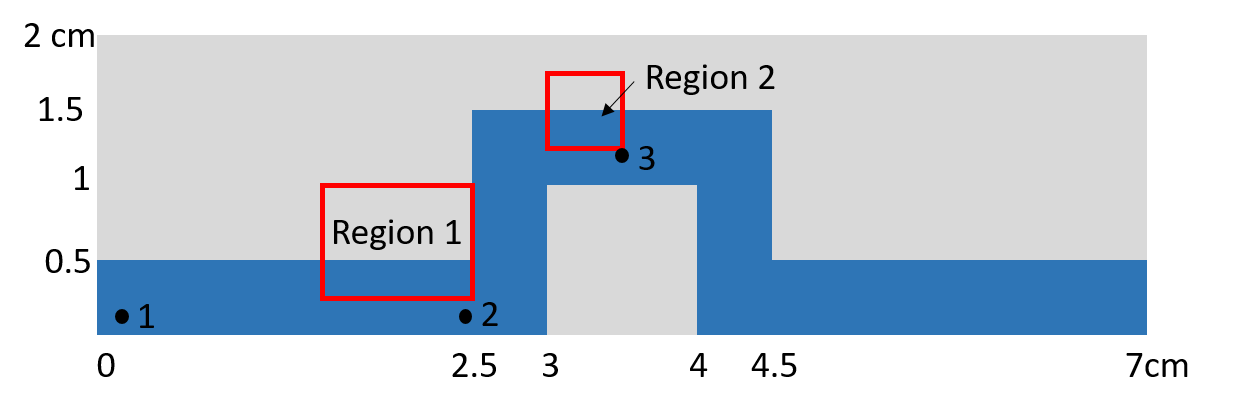}
	\caption{Crooked pipe geometry with regions 1 and 2 noted. The three points are locations for the time evolution figures. The pipe is symmetric about the bottom horizontal axis.}
	\label{fig:v7}
\end{figure}

Sensitivity analyses are carried out for a 2-D cylindrical crooked pipe problem~\cite{Graziani}. The geometry is defined in Figure \ref{fig:v7}; a radiation source turned on at $t=0$ is located at the left inlet of the optically thin material. The parameters for this problem are summarized in Table \ref{table:t1}. The problem defined by Graziani~\cite{Graziani} is modified by replacing the incoming scalar intensity boundary condition at the left inlet with an open boundary condition and a volumetric source. The source strength is 0.1 GJ/cm$^3$ns distributed throughout the first two cells along the horizontal axis and from r=0 cm to r=0.5 cm. 

\begin{table}[H]
	\begin{center}
		\caption{Parameter values for cylindrical crooked pipe example.}
		\label{table:t1}
		\hspace{1cm}
		\begin{tabular}{|c|c|}
			\hline
			Parameter & Value \\
			\hline
			dr=dz     & 0.05 (cm)\\
			dt         & 0.01 (ns)  \\
			$\rho$ (thin)  & 0.01 (g/cm$^3$) \\
			$\rho$ (thick)  & 10 (g/cm$^3$) \\
			C$_{\mathrm{v}}$ (thin)  & 0.05 (GJ/g/keV) \\
			C$_{\mathrm{v}}$ (thick)  & 0.05 (GJ/g/keV) \\
			$\kappa$    & 20 (cm$^{2}$/g) \\
			T$_{\mathrm{initial}}$   & 0.005 (keV) \\
			S   & 0.1 (GJ/cm$^3$ns) \\
			\hline
		\end{tabular}
	\end{center}
\end{table}
Figure \ref{fig:v7} illustrates the pipe geometry with two regions of interest indicated for the sensitivity analysis.

\begin{figure}[h] 
	\centering
	\includegraphics[width=1.0\textwidth]{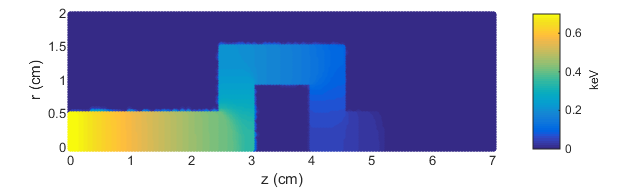}
	\caption{Radiation temperature at 10 ns for temperature independent opacity.}
	\label{fig:v5}
\end{figure}

\begin{figure}[h] 
	\centering
	\includegraphics[width=1.0\textwidth]{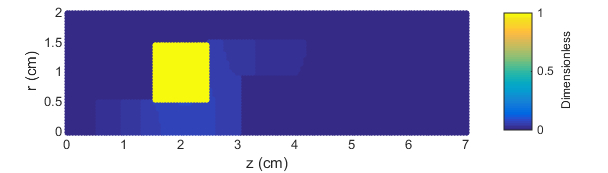}
	\caption{Adjoint scalar intensity at 10 ns. The quantity of interest is the absorption in region 1, indicated on Figure 1.}
	\label{fig:v6}
\end{figure}

\begin{figure}[h] 
	\centering
	\includegraphics[width=0.6\textwidth]{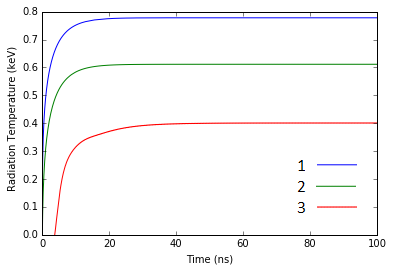}
	\caption{Time evolution of the radiation temperature at locations indicated in Figure 1.}
	\label{fig:v5a}
\end{figure}

\begin{figure}[h] 
	\centering
	\includegraphics[width=0.6\textwidth]{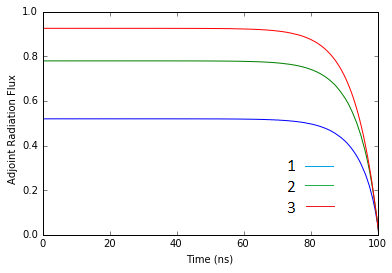}
	\caption{Time evolution of the adjoint scalar intensity at locations indicated in Figure 1. The quantity of interest is the total absorption in the system.}
	\label{fig:v6a} 
\end{figure}

We first consider the case of constant opacity. Under this assumption the last three terms in the adjoint temperature equation, givne in equation~\eqref{8}, vanish.
For illustrative purposes, Figures \ref{fig:v5} and \ref{fig:v6}  depict the forward and adjoint scalar intensity at 10 ns. The forward solution is plotted in terms of radiation temperature, defined by $\phi$$=acT$$_R^4$. Figures \ref{fig:v5a} and \ref{fig:v6a} illustrate the time dependence of the solutions at the three points numbered in Fig. \ref{fig:v7}.

 The quantities of interest are the energy absorption within the entire system and within regions 1 and 2, indicated in Figure \ref{fig:v7}. The perturbed parameters include the source, opacity, densities, and the exponent $n$ in the diffusion coefficient. The adjoint approach is compared to the traditional perturbation theory result; the method in which the problem is solved twice to compute a single sensitivity. The resulting sensitivities are summarized in Table \ref{table:t2}. In the table, the ``forward" method refers to the perturbation result, in which the forward system is solved twice and the sensitivity is computed as a finite difference. The ``adjoint" method refers to the adjoint sensitivity. The absorption is computed by integrating $\phi$$\cdot$ $\kappa$$\rho$ over the region and time period of interest.

\begin{table}
	\begin{center}
		\caption{Forward and adjoint sensitivities of energy absorption with temperature-independent opacity, integrated over the first 10 ns. }
		\label{table:t2}
		\hspace{1cm}
		Total volume; absorption= 0.59596 GJ
		\begin{tabular}{|c|c|c|c|}
			\hline
			$\theta$ & Forward Sensitivity& Adjoint Sensitivity & \% Difference \\ \hline
			$\kappa$  & 2.4629 & 2.5417 & 3.10 \\
			$\rho$$_{\mathrm{thin}}$  & 2.4629 & 2.3952 & 2.75 \\
			$\rho$$_{\mathrm{thick}}$  & -0.0012 & -0.00118 & 1.67 \\
			S  & 5.9533  & 5.8951  & 0.98    \\
			n  & 0.1353  & 0.1369 & 1.17 \\
			\hline
	    	\end{tabular}		\\
		    \vspace{1mm}
		    Region 1; absorption = 0.00487 GJ
		    \begin{tabular}{|c|c|c|c|}				
			\hline
			$\theta$ & Forward Sensitivity& Adjoint Sensitivity & \% Difference \\ \hline
			$\kappa$  &-0.0844  & -0.0871  & 3.09 \\
			$\rho$$_{\mathrm{thin}}$ & -0.000752 & -0.00073 & 2.93  \\
			$\rho$$_{\mathrm{thick}}$ & -0.08434 & -0.08673 & 2.76  \\
			S  & 0.04407  & 0.0445  & 0.97  \\
			n  & -0.01339  &  -0.01359  & 1.47 \\
			\hline
	    	\end{tabular}		\\
	    	\vspace{1mm}
	    	Region 2; absorption = 0.00172 GJ
	    	\begin{tabular}{|c|c|c|c|}				
	    	\hline
			$\theta$ & Forward Sensitivity& Adjoint Sensitivity & \% Difference \\ \hline
			$\kappa$  &-0.0595  &-0.0582  & 2.18  \\
			$\rho$$_{\mathrm{thin}}$  & -0.0603 &-0.0585 & 2.99  \\
			$\rho$$_{\mathrm{thick}}$  & -0.00077 & -0.00075 & 2.60 \\
			S  & 0.0162  & 0.0165  & 1.82 \\
			n  & 0.0000497  &  0.0000492 & 1.01\\
			\hline
		\end{tabular}
	\end{center}
\end{table}

The direct and adjoint methods give the same order of importance for the sensitivities, and differ in magnitude by less than 5\%. Thus, the finite difference approximations to the derivatives of the QOI do not introduce a significant amount of error; mesh refinement could further improve the approximation if necessary. 
The adjoint method is only advantageous if it requires less computational time than the perturbation approach. The time to complete the forward solve is 31.6 s; it would take 63.2 s to compute a single sensitivity using the perturbation method. The time to complete the adjoint solve is 31.1 s, with an additional 2.8 s to compute the sensitivities. The adjoint approach produces five sensitivities in approximately half the time required for the perturbation method to compute a single sensitivity. 

To illustrate the advantage of the adjoint approach for a more realistic system, temperature dependent material properties are considered. The crooked pipe problem is solved again with a temperature dependent opacity~\cite{Lane2013}~\cite{Wooten}:
\begin{equation}
\kappa=\frac{\kappa_0}{T^3},
\end{equation}
where $\kappa$$_0$ is the value in Table 1. The resulting sensitivities are summarized in Table \ref{table:t7}. 

\begin{table}
	\begin{center}
		\caption{Forward and adjoint sensitivities of energy absorption with temperature dependent opacity, integrated over the first 10 ns. }
		\label{table:t7}
		\hspace{1cm}
			Total volume; absorption = 269.4 GJ   
			\begin{tabular}{|c|c|c|c|}
			\hline
			$\theta$ & Forward Sensitivity& Adjoint Sensitivity & \% Difference \\ \hline			
			$\kappa$  & 1.167 & 1.225 & 4.73  \\
			$\rho$$_{\mathrm{thin}}$  & 5.172 &  5.427 & 4.70 \\
			$\rho$$_{\mathrm{thick}}$  & 1.625 & 1.584 & 2.52 \\
			S  & 5.8557  & 5.9534  & 1.64    \\
			n  & -0.0377  & -0.0381 & 1.05 \\ \hline
	    	\end{tabular}		\\
	    	\vspace{1mm}
			Region 1; absorption = 20.87 GJ
			\begin{tabular}{|c|c|c|c|}				
			\hline
			$\theta$ & Forward Sensitivity& Adjoint Sensitivity & \% Difference \\ \hline			
			$\kappa$  & 0.10455 & 0.101  & 3.518 \\
			$\rho$$_{\mathrm{thin}}$ & $<$ 10$^{-10}$ & $<$ 10$^{-10}$ & 0  \\
			$\rho$$_{\mathrm{thick}}$ & 0.01045 & 0.0108 & 3.24 \\
			S  & $<$ 10$^{-10}$ & $<$ 10$^{-10}$ & 0  \\
			n  &  0.0000206 & 0.0000213  & 3.29 \\ \hline
		    \end{tabular} \\
			\vspace{1mm}		
			Region 2; absorption = 7.215 GJ   \\				\begin{tabular}{|c|c|c|c|}				
			\hline	
			$\theta$ & Forward Sensitivity& Adjoint Sensitivity & \% Difference \\ \hline						
			$\kappa$  & 0.03612 & 0.03687 & 2.03  \\
			$\rho$$_{\mathrm{thin}}$  & 1.0*10$^{-7}$ & 1.044*10$^{-7}$ & 4.21  \\
			$\rho$$_{\mathrm{thick}}$  & 0.0361 & 0.0367 & 1.63 \\
			S  & 0.0000197  & 0.00002  & 1.79  \\
			n  &  -1.5*10$^{-7}$ & -1.548*10$^{-7}$  & 3.23 \\						
			\hline
		\end{tabular}
	\end{center}
\end{table}

In this more complicated problem the adjoint method quantifies the sensitivities with less than 5\% error and predicts the correct order of importance for the parameters. With temperature dependent opacities, the forward system takes 37.03 s to solve; thus a single sensitivity computed via the perturbation method can be obtained in 74.06 s. The adjoint temperature is no longer zero, thus the set of coupled adjoint equations is solved; this task takes 31.8 s, and the five inner products for the sensitivities take 2.9 s to compute. Even in the more complicated case of temperature dependent opacity, multiple sensitivities can be computed in less computational time than a single sensitivity can be calculated via the perturbation approach. 

\section{Conclusions}
The adjoint equations for a flux-limited radiative diffusion model are derived and used to perform sensitivity analyses. Due to the nonlinearity of the diffusion coefficient, the sensitivity expressions contain cumbersome derivatives; proposed finite difference approximations to these derivatives are shown to not introduce significant error. The adjoint method is particularly advantageous when the sensitivity of a quantity of interest with respect to multiple variables is needed. For the crooked pipe example, the forward system is composed of two coupled nonlinear partial differential equations. Solving the forward system iteratively requires more computational time than the adjoint equations, which are a set of linear coupled equations. The complete adjoint analysis for five sensitivities requires less computational time than two forward solves required to compute a single finite difference sensitivity, and quantifies the sensitivities without significant error.
The methodology presented in this work could readily be applied to transport and radiation hydrodynamics. The adjoint equations can be derived using the Lagrangian approach presented; it is expected that the adjoint method will be computationally efficient when two or more sensitivities are of interest. The transport equations are more complicated to solve, but allow for easier incorporation of non-homogeneous boundary conditions to consider a wider variety of realistic problems.

\section{Acknowledgment}
This material is based upon work supported by the Department of Energy, National Nuclear Security Administration, under Award Number DE-NA0002376.

\section*{References}
\bibliography{mybibfile}

\end{document}